\documentclass[preprint,authoryear,5p,times,twocolumn]{elsarticle}

 \usepackage{graphics}
 \usepackage{epsfig}

\usepackage{amssymb}

\def\apj{{\em Astrophys.~J.}}
\def\apjl{{\em Astrophys.~J.~Lett.}}
\def\apjs{{\em Astrophys.~J.~Suppl.}}
\def\araa{{\em Ann.~Rev.~Astr.~Astrophys.}}
\def\mnras{{\em Mon.~Not.~R.~astr.~Soc.}}
\def\nat{{\em Nature}}
\def\pasj{{\em Publ.~astr.~Soc.~Japan}}
\def\pasp{{\em Publ.~astr.~Soc.~Pacif.}}

\newcommand{\etal}{{et al.}}

\newcommand{\Msun}{\>{\rm M_{\odot}}}

\journal{Proceedings SCSLSA-7}

\begin{document}

\begin{frontmatter}



\title{Emission Lines as a Tool in Search for Supermassive Black Hole
Binaries and Recoiling Black Holes}


\author{Tamara Bogdanovi\'c}
\ead{tamarab@astro.umd.edu}

\address{Department of Astronomy, University of Maryland, College
Park, MD 20742}

\author{Michael Eracleous}

\author{Steinn Sigurdsson}

\address{Department of Astronomy \& Astrophysics and Center for
Gravitational Wave Physics, Pennsylvania State University, State
College, PA 16802}

\begin{abstract}
 
Detection of electromagnetic (EM) counterparts of pre-coalescence
binaries has very important implications for our understanding of the
evolution of these systems as well as the associated accretion
physics.  In addition, a combination of EM and gravitational wave
signatures observed from coalescing supermassive black hole binaries
(SBHBs) would provide independent measurements of redshift and
luminosity distance, thus allowing for high precision cosmological
measurements. However, a statistically significant sample of these
objects is yet to be attained and finding them observationally has
proven to be a difficult task. Here we discuss existing observational
evidence and how further advancements in the theoretical understanding
of observational signatures of SBHBs before and after the coalescence
can help in future searches.

\end{abstract}


\end{frontmatter}



\section{Introduction\label{observations}}

The most direct evidence for the formation of supermassive black hole
binaries (SBHBs) currently available through electromagnetic (EM)
observations is the existence of black hole pairs spatially resolved
on the sky. An example of such a system is NGC~6240, an ultra-luminous
infrared galaxy in which the $Chandra$ X-ray observatory revealed a
merging pair of X-ray emitting active nuclei with a separation of
$\sim$1 kpc \citep{komossa}. The large observed separation implies
that the two supermassive black holes (SBHs) are a pair of unrelated
massive objects orbiting in the potential of the host galaxy that are
yet to reach the binary stage, in which the two SBHs will be bound by
their mutual gravitational force. Practical obstacles in the direct
identification of close binaries via EM observations arise from the
need for a very high spatial resolution and accuracy in position
measurements. Because of the small orbital separation of sub-parsec
binaries and the small separation of recoiling SBHs from the centers
of their host galaxies not too long after the coalescence\footnote{SBH
recoil considered here is due to postulated asymmetric gravitational
wave emission, as shown by numerical relativity.}, the best prospect
for {\it direct imaging} of these objects is offered by radio
interferometers, provided that SBHBs and recoiling SBHs are radio
sources. Since this method relies on pointed observations of candidate
objects, it requires {\it a priori} knowledge about their location on
the sky. This technique yielded one binary candidate, 0402+379, an
object with two compact, unresolved radio-cores at a projected
separation of only $\sim7$~pc, observed with the {\it Very Long
Baseline Array} \citep[{\it VLBA};][]{maness04,rodriguez06}. Followup
multi-wavelength observations of this and similar radio candidates in
the future are essential for studying the rest of associated
observational signatures.

The second most compelling piece of observational evidence for a SBHB
is a periodic signal associated with Keplerian motion, such as the
periodic or quasi-periodic variability of the light emitted by an
object.  A binary candidate selected according to this criterion is
the blazar OJ~287, which exhibits outburst activity in its optical
light curve with a period close to 12 years \citep{valtonen08}.
Distinguishing periodic variability due to a binary from the common
variability of active galactic nuclei (AGNs) can be challenging
because it relies on long-term monitoring of the candidates and is, in
many cases, less pronounced than in OJ~287.

\begin{figure}[t]
\center{
\includegraphics[width=0.37\textwidth]{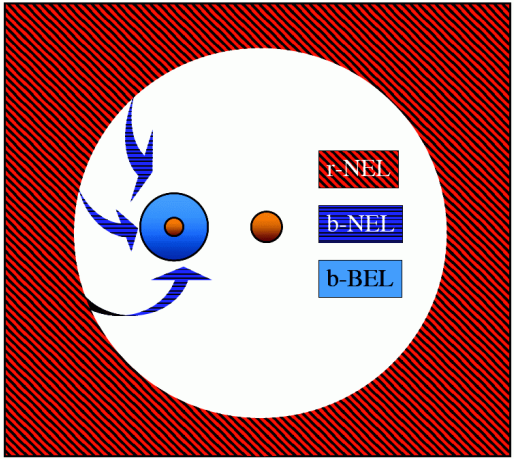}}
\caption[j0927]{Illustration of the binary model for J0927.  The
  innermost region of a circumbinary disk is shown after the binary
  has cleared a low density ``hole'' in the disk (top view, not drawn
  to scale). In this model r-NELs are associated with the circumbinary
  disk, b-BELs with the disk surrounding the less massive secondary
  SBH, and b-NELs with the accretion streams flowing from the inner
  edge of the circumbinary disk toward the secondary. Figure adapted
  from \citet{bogdanovic09}. \label{fig1}}
\end{figure}

\begin{figure}[t]
\center{
\includegraphics[width=0.35\textwidth]{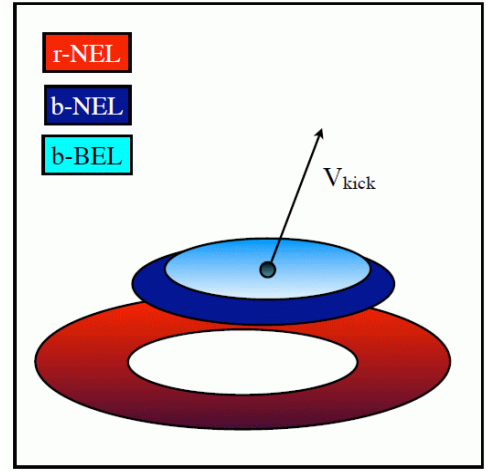}}
\caption[j0927]{Illustration of the model of a recoiling SBH proposed
for J0927 by \citet{komossa08b} (oblique view, not drawn to scale). In
this model r-NELs are associated with the ISM of the host galaxy,
b-BELs with the disk recoiling with the SBH, and b-NELs with the
expanding accretion disk, swept up ISM, or associated
outflows. \label{fig2}}
\end{figure}

The third technique used to select SBHB candidates relies on a {\it
detection of the Doppler-shift} in the spectrum of an object that
arises from the orbital motion of a binary. In this case the search
utilizes the emission-lines associated with multiple velocity systems
in the spectrum of a candidate object. Because this effect is also
expected to arise in case of a recoiling SBH receding from its host
galaxy, the same approach is used to ``flag'' candidates of that type.
The Doppler shift signature is however not unique to these two
physical scenarios, and complementary observations are needed in order
to determine the true nature of observed candidates (as discussed in
\S\ref{candidates}).  The advantage of the method is its simplicity,
as emission spectra that exhibit Doppler shift signatures are
relatively straightforward to select from large archival data sets,
such as the Sloan Digital Sky Survey (SDSS). So far a few SBHB
candidates have been selected out of $\sim21,000$ SDSS quasars:
J092712.65+294344.0 \citep[hereafter J0927;][]{bogdanovic09, dotti09},
J153636.22+044127.0 \citep[J1536;][]{bl09}, and J105041.35+345631.3
\citep[J1050;][]{shields09}, where the first and the last object have
also been flagged as recoiling SBH candidates
\citep{komossa08b,shields09}. In the following section we describe in
more detail results and uncertainties associated with this approach,
using J0927 as an illustrative example, and summarize the status of
the other two candidates.

\subsection{SBHB candidates: J0927, J1536, and J1050\label{candidates}}

The optical spectrum of J0927 features two sets of emission lines
offset by $2650\,{\rm km\,s^{-1}}$ with respect to each other. The
``redward" system consists of only narrow emission lines (r-NELs),
while the ``blueward" emission-line system comprises broad Balmer
lines (b-BEL) and narrow, high-ionization forbidden lines (b-NELs).
In the context of the binary model, the observed velocity shift
represents the projected orbital velocity of the less massive member
of a bound black hole pair. In this model, the emission lines of the
blueward system (b-NELs and b-BELs) originate in gas associated with
the less massive, secondary black hole, while the r-NELs originate in
the interstellar matter (ISM) of the host galaxy. An illustration of
the proposed geometry for the binary model is shown in Figure~1. For
an assumed inclination and orbital phase angles of 45$^\circ$ and a
mass ratio of $q=0.1$, the inferred orbital period and mass of the
binary are $\sim 190$~yr and $\sim 10^9\,\Msun$, respectively. The
accreting secondary is the main source of ionizing radiation while the
primary is either quiescent or much fainter.

According to the recoiling SBH interpretation for J0927 \citep[note
that this was the first explanation proposed for this object
by][]{komossa08b}, the b-BELs originate in the broad-line region
retained by the SBH, the b-NELs were attributed to gas that is
marginally bound to the SBH, swept up or outflowing, and the r-NELs
are associated with the ISM of the host galaxy (the geometry of the
recoiling SBH model is outlined in Figure~2). In this model the
recoiling SBH of mass $\sim 6\times10^8\,\Msun$ appears as an AGN
during the period in which it accretes from the disk that is carried
along with the hole. The accreting SBH provides a source of ionizing
radiation for the b-BELs and b-NELs, as well as for the r-NELs, albeit
from a larger distance.

Another class of proposed explanations for J0927 includes interaction
of two galaxies in the potential well of a massive cluster
\citep{heckman09} and a superposition of otherwise unrelated AGN in a
cluster environment \citep{sbs09}. Both of these scenarios can, in
principle, explain the multiple velocity-emission-line systems seen in
the spectrum of J0927 and offer a good alternative to the SBHB and
recoiling SBH models. In a subsequent paper, however,
\citet{decarli09a} reported that J0927 is unlikely to be a member of a
rich cluster. At this point every model proposed for J0927 received
some amount of criticism (see the above papers for details) and thus,
the nature of J0927 is not conclusively determined and requires
further investigation.

Soon after its discovery, the interpretation for the candidate J1536
was called into question because of the lack of the variability
expected from Keplerian motion \citep{chornock09a, gaskell09} and the
similarity of its line profiles to those of ``double-peaked emitters''
\citep{chornock09b}. At this point, the data available for this
object do not conclusively rule out the binary scenario and further
observational tests have been suggested \citep{wl09,lb09,decarli09b}.

Most recently, \citet{shields09} reported the discovery of another
unusual SDSS quasar, J1050, for which an SBHB is a viable explanation,
along with a recoiling SBH, and a double-peaked emitter. As with the
previous cases, further observations are needed in order to
discriminate among proposed hypotheses.

J0927, J1536, and J1050 do not make up a uniform class of objects
because their emission-line spectra differ from each other.  A common
property shared by all three objects is that their emission line
systems exhibit shifts in the range $2000-4000\,{\rm km\,s^{-1}}$. If
such velocity shifts are interpreted in the context of a binary model,
and if the emission-line systems are associated with one or both of
the black holes, the expectation is that the orbital motion of the
binary should give rise to a measurable velocity change of the
emission lines on a time-scale of several years. For all three objects
this basic test was carried out by taking spectra of the object at a
few different epochs and measuring the change in the position of the
emission line peaks. In all three cases the measured rate of velocity
shift is very low ($dv/dt\sim{\rm few}\times10\,{\rm
km\,s^{-1}\,yr^{-1}}$) and consistent with zero within the error
bars. Taken at the face value this result seems to eliminate the
possibility that any of the three candidate objects is a
binary. However, it may also be that if the selection process is
biased towards binaries with large velocity shifts, they may be
easiest to spot precisely when they are close to quadrature and
$dv/dt$ is small. Hence, it may take longer to detect a change in $dv$
than expected if time scale is estimated based on a random phase
prior. Moreover, the uncertainties in emission geometry and radiative
processes of binary accretion flows currently preclude elimination of
the binary model based on the lack of velocity shifts measured from
the line peaks (see discussion in \S\ref{transfer}).

In the following section we briefly describe theoretical ideas for
formation and evolution of SBHBs, including the consequences of their
coalescence.

\section{Astrophysics of SBHBs before and after coalescence\label{theory}}

Galactic mergers are expected to be the major route for formation of
binary black holes, in agreement with predictions of hierarchical
models of structure formation \citep*{haehnelt02, volonteri} and the
observation that the majority of galaxies harbor massive black holes
in their centers \citep*[e.g.,][]{kr, richstone, peterson00,
ff05}. Following a galactic merger, the two SBHs are initially carried
within their host bulges and eventually find themselves at $\sim$kpc
separations \citep[e.g.,][]{kazantzidis05} from each other. In this
first evolutionary stage, the separation of a pair of the SBHs
decreases through the process of dynamical friction that involves
two-body gravitational interaction of individual SBHs with stars and
interactions with gas in the nuclear region. In the second stage (at
separations $\lesssim$10~pc), the binary becomes gravitationally bound
when the binary mass exceeds that of the gas and stars enclosed within
its orbit. The evolution of the binary in this phase is determined by
the availability of stars for three-body interactions
\citep{berczik06, sesana07, perets07} and the availability of gas
which can transport the excess angular momentum of the binary
\citep{bbr, ivanov99, gr00, armitage02, escala1, escala2, dotti06a,
dotti07, mayer07, colpi07, cuadra08, hayasaki08}. In the third and
final stage (at separations $<10^{-2}$~pc), the binary is led to
coalesce by the emission of gravitational waves (GWs) and forms a
single, remnant SBH with mass slightly lower (typically by several
percent) than that of the two parent SBHs combined. Due to asymmetries
in the orientation of the black hole spin axes with respect to the
binary orbital axis, and the masses of the two SBHs with respect to
each other, the emission of GWs is, in general, not isotropic.  The
GWs carry net linear momentum in some direction, causing the center of
mass of the binary to recoil in the opposite direction.

If the candidates described in \S\ref{candidates} turn out to be
SBHBs, these would most likely correspond to binaries in the second
evolutionary stage. Several physical processes may contribute to the
presence of a, hopefully unique, spectral signature of such binaries:
the presence of one or two (unresolved) accretion disks {\it around
individual black holes}, albeit in general of different luminosity;
the tidal distortion of the inner parts of the {\it circumbinary} disk
by the presence of the binary (similar to features seen in
proto-planetary disks).

The ``mass loss'' and gravitational rocket effect at the end of the
third evolutionary stage of the binary can have profound effects on
the surrounding gas and potentially give rise to unique observational
signatures as was recently suggested by a number of authors
\citep{loeb07, sb08, lippai08, sk08, kl08, moh08, devecchi08,
  oneill09}. Namely, the results of numerical relativity calculations
show that recoil velocity of a remnant SBHs can range from about 200
${\rm km\,s^{-1}}$ for SBHs with low spins or spin axes aligned with
the orbital axis \citep{herrmann07a, herrmann07b, baker06, baker07,
  gonzalez07b} to a remarkable $\sim$4000 ${\rm km\,s^{-1}}$ in case
of maximally spinning SBHs with spins oppositely directed and in the
orbital plane \citep{campanelli07a,campanelli07b}. If following a
galactic merger the two SBHs find themselves in a gas rich
environment, gas accretion torques will act to align their spin axes
with the orbital axis, thus reducing the maximum kick to a value well
below the galactic escape speed \citep[the escape speed from most
galaxies is $< $2000 ${\rm km\,s^{-1}}$,][]{merritt04}. This effect is
expected to increase the chance of retention of recoiling SBHs by
their host galaxies \citep{bogdanovic07}, in agreement with the
observation that almost all galaxies with bulges appear to have a
central SBH \citep{ff05}.  However, there may be some fraction of
remnant SBHs formed in the aftermath of gas poor mergers, where gas
torques are insufficient to prevent the high velocity recoil,
launching the SBH out of the potential well of a host bulge or a host
galaxy. One unique spectral signature of such systems could result
from the distortion of a nuclear accretion disk and due to the
velocity shift of the portion of the disk that is gravitationally
bound to the SBH. On the basis of this signature, \citet{komossa08b}
reported a discovery of the first recoiling SBH candidate, J0927,
followed by J1050, reported by \citet{shields09}, (see
\S\ref{candidates}).

\begin{figure}[t]
\center{
\includegraphics[width=0.35\textwidth]{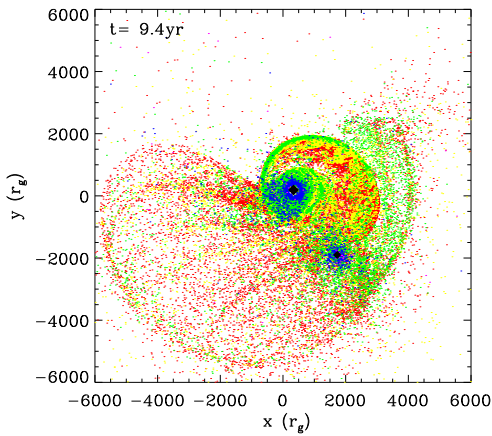}}
\caption{Snapshot from a simulation showing a SBHB and gas projected
  onto the plane of the binary orbit in the co-rotating model at 9.4~yr
  after the beginning of the simulation.  The rotation of the binary
  and the disk is counter-clockwise.  The temperatures of gas
  particles are marked with color: red $T < 10^4$~K; yellow $10^4 {\rm
    K} < T< 10^6$~K; green $10^6 {\rm K}< T < 10^8$~K; and blue $10^8
  {\rm K} < T < 10^{10}$~K. The higher temperature particles are
  plotted over the lower temperature ones, with the result that some
  information is hidden. Figure~4 illustrates this effect and can also
  be used as a color bar. Figure adapted from
  \citet{bogdanovic08}. \label{fig3}}
\end{figure}

\begin{figure} [t]
\includegraphics[width=0.49\textwidth]{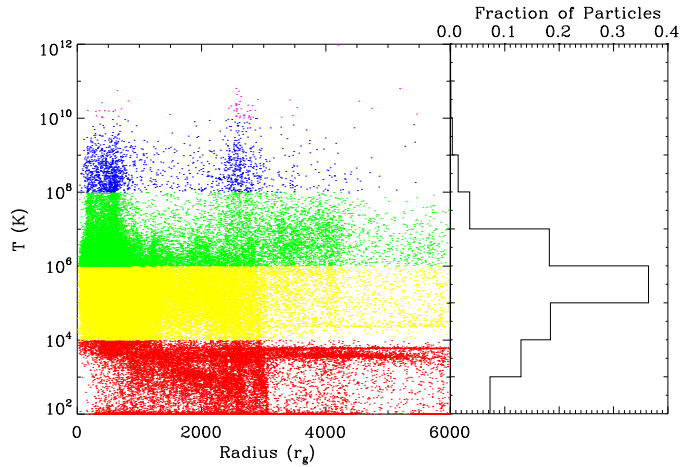}
\caption{{\it Left:} Temperatures of gas particles as a function of
  radius at 9.4 years in the co-rotating model simulation. This figure
  shows the extent of temperature stratification in the
  disk. Different colors that mark each temperature band are
  equivalent to those in the previous figure. This figure corresponds
  to the morphology of the disk shown in Figure~3. {\it Right}:
  Histogram showing the temperature distribution of particles.  Figure
  adapted from \citet{bogdanovic08}. \label{fig4}}
\end{figure}

\begin{figure}[t]
\center{
\includegraphics[width=0.37\textwidth]{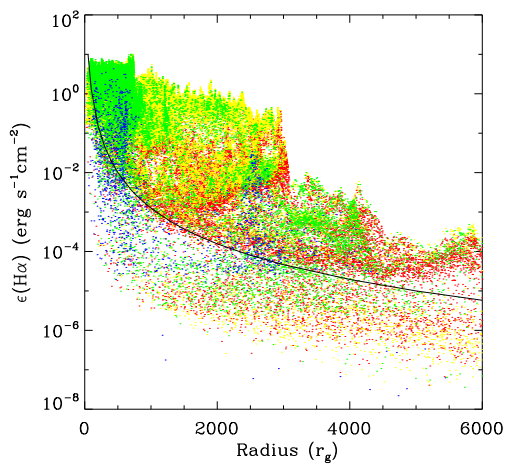}}
\caption{H$\alpha$ emissivity of the gas as a function of radius, 
at 9.4 years for a co-rotating model with solar metallicity
gas. The emissivity of each gas cell plotted in the figure is weighted
by the density of the gas at that position, so that comparison with
the parametric emissivity model (plotted as a solid line with
arbitrary normalization) can be made.  The color legend is the same as
in previous two figures.  Figure adapted from
\citet{bogdanovic08}. \label{fig5}}
\end{figure}

\begin{figure}[t]
\includegraphics[width=0.45\textwidth]{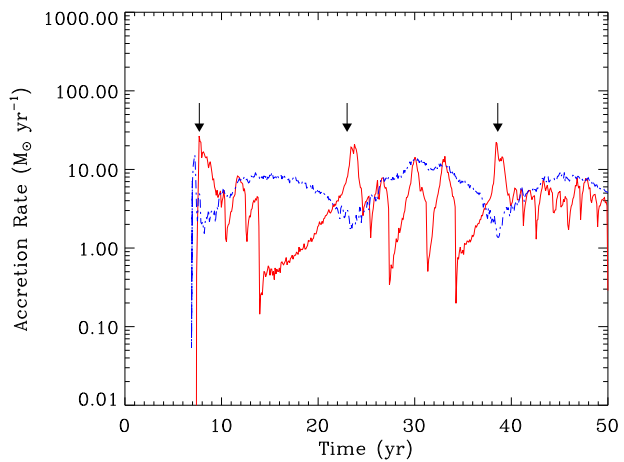}
\caption{Effective accretion rate on the primary ({\it solid, red
line}) and secondary ({\it dashed, blue line}) black hole calculated
from co-rotating model. Note that in this figure only we show data
from a lower resolution test simulation (20k particles) which ran over
a longer period of time ($\sim50$~yr) than the baseline 100k simulation
($\sim35$~yr) but is in all other aspects equivalent to it.  The
accretion rates can be translated into UV/X-ray luminosity by assuming
$1\Msun\,{\rm yr^{-1}} \approx 10^{43}\,{\rm erg\,s^{-1}}$ in the
UV/X-ray band. The arrows mark the times of the pericentric passages
of the binary.  Figure adapted from \citet{bogdanovic08}.\label{fig6}
}
\end{figure}

Therefore, the emission-line Doppler shift has so far been 
a key signature, sought after in screenings for both the SBHB and
recoiling SBH candidates. In order to utilize it to its full
potential, and in anticipation of even reacher data sets in the
future, expanded towards higher redshifts and lower-luminosity AGN, it
is necessary to gain understanding of the remainder of associated
emission signatures in other wavelength bands. An opportunity to
study observational signatures in parallel with observations is
offered by hydrodynamics simulations of SBHs interacting with gas. In
the following section we describe some of the advancements and
challenges associated with this approach in general and then present
selected results from our work.

\section{Simulations of SBHs and gas}

Over the past several years simulations of merging galaxies have
significantly contributed to our understanding of the formation and
evolution of black hole pairs (as discussed at the beginning of
\S\ref{theory}). However, simulations that would span without gaps the
entire dynamical range from a galactic merger to binary coalescence
are still computationally prohibitively expensive. In order to reduce
this problem into one that is numerically tractable it is currently
necessary to make simplifications in the initial conditions and the
treatment of the thermodynamics of the gas. These choices include
black hole pairs initialized in steady-state galactic and nuclear
accretion disks, while in reality they are most likely to be perturbed
and possibly heavily irradiated by one or two AGNs. The treatment of
gas thermodynamics in such disks may involve a parametric prescription
in the form of a fixed equation of state or a cooling law parametrized
in terms of the dynamical time scale of a system ($\beta$-model).
These simplifying assumptions result in uncertainties in the evolution
of the SBHBs with the smallest (sub-parsec) orbital separations; such
binaries are also the most interesting because of their proximity to
gravitational wave regime and coalescence. Further improvements in
the treatment of gas thermodynamics can be made thus reducing one of
the two dominant sources of uncertainty (the other one being the
initial conditions). This can be achieved by adopting a more
physically motivated form for gas cooling and also gas heating by
introducing the sources of ionization associated with the accreting
SBHs.
 
Given the challenges associated with the detection of SBHBs and
recoiling SBHs, predictions of observational signatures based on
existing hydrodynamical simulations would be very valuable.  This is
currently not attainable because hydrodynamical models do not include
radiative transfer prescriptions that allow such predictions to be
made (the exception are the accretion rate curves, which are also
subject to the uncertainties described in the previous paragraph).
Indeed, a complete treatment of radiative transfer within a
hydrodynamical simulation of SBHs and non-zero metallicity gas is
currently not feasible. Nevertheless, a significant step toward this
goal can be made with the help of photoionization codes that include
full radiative transfer treatment of gas composed of multiple atomic
species and can make a variety of predictions in the form of
characteristic emission lines and continua, associated optical depths,
and a plethora of other information. Photoionization codes can be used
to characterize radiative transfer within every individual parcel of
gas in a simulation, given the physical properties at the location of
a parcel (number of ionizing photons, temperature, and density of the
gas). Because the range of physical properties in a simulation can be
predicted with some certainty (based on the properties of nuclear
regions in galaxies hosting AGNs or from test simulations), the
response of the gas to photoionization {\it precomputed} and stored in
a database. Such a database can be used for reference during
simulations \citep[for an example of this approach
see][]{bogdanovic08}, or in some cases in the post-processing of
simulation data \citep{bogdanovic04}, in order to calculate
observational signatures. The main effort associated with this
approach is the calculation of well behaved heating, cooling, and
emissivity maps (within the limits of the specific photoionization
code). The computational cost associated with the addition of the
photoionization data grid to the hydrodynamical code is modest, as the
majority of the additional operations are interpolations between the
grid points. While any such method must acknowledge that detailed
modeling of emission line and other spectral properties is currently
still out of reach, it still offers the best means of examining the
most {\it characteristic} signatures of sub-parsec SBHBs and recoiling
SBHs.

\begin{figure}[t]
\hspace{-4mm}
\includegraphics[width=0.51\textwidth]{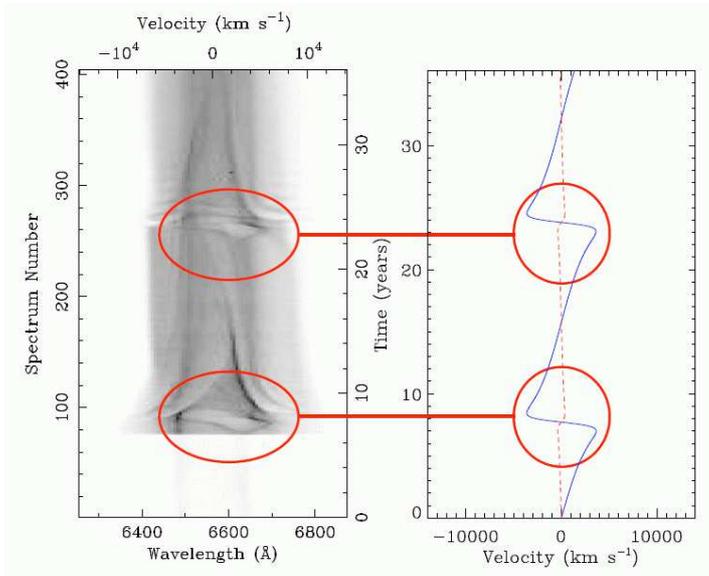}
\caption[prof]{Trailed spectrogram (left) and black hole projected
velocity curves (right) plotted for the co-rotating model. Trailed
spectrogram is a logarithmic gray scale map of the H$\alpha$ intensity
against wavelength and observed velocity. Darker shades mark higher
intensity. Notice the low relative intensity of the profiles before
the start of accretion. The velocity curve panel shows the orbital
velocities of the primary ({\it dashed, red line}) and secondary ({\it
solid, blue line}) black holes projected along the line of sight to
the observer.  The velocity curves are skewed because of the non-zero
eccentricity of the orbit.  Characteristic features associated with
the pericentric passages of the binary in years 8 and 23 are marked on
both figures. Figure adapted from \citet{bogdanovic08}.\label{fig7}}
\end{figure}

\begin{figure}[t]
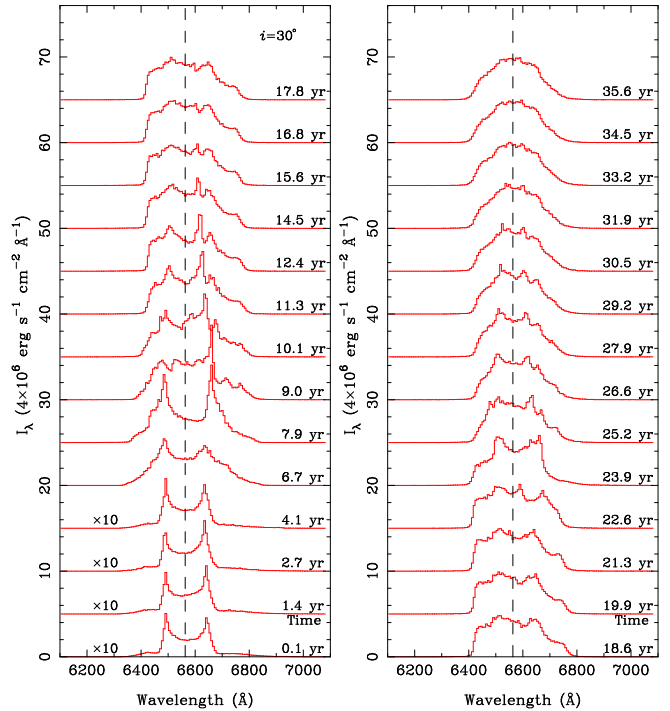

\includegraphics[width=0.23\textwidth]{prof1.ps}
\includegraphics[width=0.23\textwidth]{prof2.ps}
\caption[prof]{Sequence of H$\alpha$ emission-line profiles selected
from a model where the disk is co-rotating with respect to the orbit
of the SBHB. The intrinsic intensity of profiles is plotted against
wavelength.  The first 4 profiles in the sequence are multiplied by a
factor of 10, so that they can be represented on the same intensity
scale with the other profiles. The corresponding time from the
beginning of the simulation is given next to each profile. The
observer is located at infinity, in the positive $xz$-plane at an
angle of $i=30^{\circ}$ with respect to the axis of the binary orbit.
Figure adapted from \citet{bogdanovic08}.\label{fig8}}
\end{figure}

In the following section we summarize the results from such a study of
observational signatures of SBHBs interacting with gas.

\subsection{Investigations of observational signatures using multi-species radiative transfer}\label{transfer}

We implemented an approximate multi-species radiative transfer scheme
in a modified version of the smoothed particle hydrodynamics (SPH)
code {\it Gadget~1} \citep{springel01} in order to study the
observational signatures of sub-parsec binaries interacting with gas
\citep{bogdanovic08}. The gas is assumed to have solar metallicity and
its physical properties were characterized by calculating heating,
cooling, and radiative processes as an integral part of hydrodynamical
simulations. Using photoionization code {\it Cloudy} \citep{ferland},
we constructed a grid of photoionization maps for a range of
parameters, determined from preliminary simulations without
cooling. The maps are calculated in the parameter space of {\it
density} and {\it temperature} of the gas and {\it intensity} of the
ionizing radiation. The range of parameter values for which the maps
were computed is as follows: $10^9\,{\rm cm^{-3}} < n < 10^{19}\,{\rm
cm^{-3}}$, $2000\,{\rm K} < T < 10^8$~K, and $0\; {\rm erg \;cm^{-2}
\, s^{-1}} < J < 10^{17}\, {\rm erg \;cm^{-2} \, s^{-1}}$. Gas at
higher temperatures than $10^8$~K is also found in our simulations; in
such cases we calculate heating and cooling rates and other properties
of interest by linearly extrapolating the grid values. In the SPH
simulations we set the lower threshold for the gas temperature to
100~K and assign no upper threshold. Based on these results we
calculated the accretion-powered continuum and H$\alpha$ light curves,
as well as the H$\alpha$ emission line profiles emerging from the
inner parts of a gas disk in the presence of a binary, on a scale of
$<0.1$ pc. We selected the Balmer series H$\alpha$ line ($\lambda_{\rm
rest} = 6563$~\AA) because it reflects the kinematics in the broad
line region of AGN \citep[e.g.,][]{sulentic00}. Hence, by mapping the
kinematics in the region of sub-parsec size, broad H$\alpha$ line can
in principle provide important information about SBHBs interacting
with gas\footnote{We do not model the narrow component of H$\alpha$
line which is thought to originate at larger distances.}. Moreover,
H$\alpha$ is the most prominent broad line in the optical spectrum,
least contaminated by neighboring narrow emission lines. The H$\alpha$
line in accretion-powered sources is thought to be powered by
illumination of a disk by an AGN and the radiative transfer associated
with it is relatively well understood \citep{csd89}.  That said, the
method can be extended to include other transitions of interest, given
that calculation of the photoionization data grid with {\it Cloudy}
includes a wide range of atomic species from hydrogen to zinc.

\subsubsection{Physical properties of the gas:}

In the remainder of this section we focus on selected results from the
simulation of the binary co-rotating with the gas disk and refer the
reader to the original publication for discussion of the
counter-rotating case.  The simulation follows the evolution of a SBHB
over 35 years in total and comprises $10^5$ SPH gas particles (100k
hereafter). The black hole mass ratio is 10:1, with the primary having
a nominal mass of $M_1=10^8 \Msun$. The secondary SBH is on an
eccentric orbit around the primary with $e=0.7$, and the initial
orbital period of the binary is close to 16 years.  The gas disk is
initially associated with the primary black hole only and its mass and
outer radius are initially $10^4 \Msun$ and $2\times 10^3 r_g$ ($r_g
\equiv GM_1/c^2$).  The median value of the initial disk density is
$\sim 10^{13}\;{\rm cm^{-3}}$.  The geometry of the system and the
temperature of the gas are illustrated in Figure~3 at $t=9.4$~yr after
the beginning of the simulation.  The primary SBH is initially located
at $(x,y,z) = (0,0,0)$ and exhibits modest orbital motion during the
simulation.

The gas in the simulation is heated by shocks that arise from
interactions with the secondary SBH and by radiation once the
accretion onto the black holes begins. As a result, a portion of the
gas is blown out of the plane of the disk and forms a hot halo that
envelopes the disk. The gas quickly departs from the uniform initial
temperature of 2000~K and assumes a wide range of temperatures, as
shown in Figure~4. Figure~4 further illustrates that multiple
temperature components can be present at a single radius and shows the
temperature distribution of the SPH particles.  The hot component of
the gas spends a significant amount of time in the temperature range
$10^4-10^8$ K and in this regime radiative cooling is dominated by
free-free emission and recombination. The highest value of the gas
temperature in the simulation, T$\sim10^{12}$ K, is reached after the
shock is formed by the secondary. On a time scale of months, the
temperature of the gas falls below $10^{10}$ K due to the combined
effects of radiative cooling and adiabatic expansion. On the other
hand, the cold component of the gas, confined to the higher density
spiral arm has a temperature in the range $10^3-10^4$ K, close to the
initial value, and cools largely through recombination radiation.  At
the end of the simulation the gas density in the disk is in the range
$10^{8}-10^{14}\; {\rm cm^{-3}}$.  The median density in the tenuous,
photoionized halo is $10^7-10^8\;{\rm cm^{-3}}$ and the minimum
density can be as low as $10^{5}\;{\rm cm^{-3}}$.

The interaction of the binary with the gas also affects the emissivity
of the gas and consequently its emission signatures. The most
important consideration in the calculation of the emitted H$\alpha$
light is the efficiency with which the gas reprocesses the incident
ionizing radiation and re-emits it in the form of H$\alpha$ photons.
This efficiency is commonly characterized by the surface emissivity of
the gas. The emissivity of the photoionized gas depends on numerous
physical parameters. Locally, it depends on the physical properties of
the gas. Globally, its spatial distribution depends on the structure
of the accretion disk.  We assess the H$\alpha$ emissivity of every
gas particle in the simulation using photoionization data grids
precomputed with {\it Cloudy}.  In Figure~5 we compare the emissivity
determined from our photoionization calculations with the parametric
model used for AGNs in which the emissivity is described as a function
of radius, $\epsilon\sim\xi^{-q}$ with $q\approx3$ \citep{csd89}. We
find a qualitative agreement between the two albeit, with a
significant amount of scatter.  This is not surprising given the
variations in surface density of the perturbed gas disk in our binary
model.

\subsubsection{Characteristic observational signatures:}

We further find that X-ray outbursts (powered by accretion onto the
two AGN {\it and} shocks) should occur during pericentric passages of
a coplanar binary as long as the nuclear region is not devoid of
gas. During pericentric passages the binary accretion rate reaches
$\sim30 {\rm \Msun\;yr^{-1}}$ (Figure~6), comparable to the Eddington
rate for this system, while during the remainder of the orbital period
the total accretion rate averages at the level of ${\rm few} {\rm
\Msun\;yr^{-1}}$. This implies luminosity of $L_{E} \approx
1.5\times10^{46}M_{8}\,{\rm erg\,s^{-1}}$ during outbursts and an
order of magnitude lower luminosity on average. Despite the 10:1 mass
ratio the two SBHs exhibit comparable accretion rates and at some
times the accretion rate onto the secondary exceeds that of the
primary SBH. This inversion happens the lower mass object has a
smaller relative velocity with respect to the gas than the primary and
can result in a higher relative luminosity of a lower mass object
\citep{al96, gr00, hayasaki07}. During this same period the integrated
(bolometric) luminosity of the disk is at a mean level of $\sim
10^{45}\,{\rm erg\,s^{-1}}$, and thus, comparable to that of the
photoionization sources.

At the mean level of accretion inferred from the simulation, the
gaseous disk should be depleted after only $10^3-10^4$ yr. In order
for the outburst activity to last over longer time scales, the
reservoir of gas in the nuclear region needs to be continually
replenished. It is also plausible, however, that repeated collisions
of the secondary black hole with the disk will completely disrupt it
and turn it into a spherical halo of hot gas. In such a case, the
accretion rate would be relatively smooth and uniform, and no outburst
would be evident. A calculation following the X-ray light curve
variability over a large number of orbits is necessary in order to
confirm that the periodicity is a long-lived signature of the binary.
Although currently not possible with the SPH method used here,
simulations of the long term evolution of the binary together with
hydrodynamics and radiative transfer may be achieved in the future.

In addition to the recurrent outbursts in the X-ray light curve the
signature of a binary is potentially discernible in the H$\alpha$
light, and more specifically, H$\alpha$ emission line profiles.  After
the start of accretion, the H$\alpha$ luminosity reaches
$10^{39}-10^{40}\;{\rm erg\, s^{-1}}$, observable out to the distance
of the Virgo Cluster (16~Mpc) and possibly up to the distance of the
Coma cluster (100~Mpc). The changes in the profile sequence are
easiest to discern in the trailed spectrogram (shown in the left panel
of Figure~7) which represents a 2D map of the H$\alpha$ intensity
against observed velocity (or wavelength). The Doppler boosting of the
blue side of the line and gravitational redshift of the red wing are
noticeable during the pericentric passages of the binary and, in
principle, can serve as indicators of the orbital period.  The width
of the H$\alpha$ profile increases after the pericentric passages of
the system, reflecting the inflow of gas towards the primary. Also,
the widening of the profile is asymmetric and slightly shifted towards
the red with respect to the pre-pericentric sequence of profiles. This
shift is a signature of the motion of the accretion disk which follows
the primary that is receding from the observer.  In Figure~7 we
compare the trailed spectrogram with the velocity curves of the two
binary components projected onto the line of sight to the observer and
highlight the instances of pericentric passages. The variations in the
profile intensities correspond to the features in the velocity curve
of the secondary in years 8 and 23, when it is moving towards the
observer with the highest projected velocity (i.e., the binary is in
quadrature).  Hence, we find the signatures of the velocity curves of
both black holes in the emission line profiles: if such features could
be discerned in the observed H$\alpha$ emission-line profiles, they
would signal the presence of a binary and potentially allow for
determination of its mass ratio.

The sequence of profiles shown in Figure~8 can be obtained if the
trailed spectrogram is sampled at selected times.  Initially, from the
unperturbed disk we observe double-peaked emission-line profiles but
they gradually depart from this shape as the perturbation propagates
through the gaseous disk. Our calculated profiles account only for the
{\it broad component} of the H$\alpha$ line; the narrow component is
not taken into account. The narrow component of H$\alpha$ is present
in the observed (real) profiles and is thought to originate in gas
outside of the nuclear region. Its presence adds a level of complexity
to the analysis and modeling of the observed H$\alpha$ emission-line
profiles, as the narrow component of the line needs to be removed
before the kinematics and geometry of the gas in the nuclear region
can be assessed from the broad component of the profile.\footnote{For
a more detailed discussion of phenomena that may affect the shape of
the H$\alpha$ emission line profiles as well as the effect of
approximations made in the calculation see \citet{bogdanovic08}.}

The irregular and variable profiles in Figure~8 are not unlike those
seen in a small percentage of AGN and in SBHB candidates found in
SDSS. In practice however, going from the detection of a binary
candidate to a positive identification of a SBHB based on only a few
epochs of observations of irregular broad Balmer lines is quite a
challenging task.  Our study indicates that the variability of broad
emission-line profiles may be subtle and reflected in the extended
wings of the profiles. In that sense an expectation that the orbital
motion of the binary can be inferred from the velocity shift of the
profile peaks may be simplistic and it may yield unreliable results.

\section{Conclusions}

Understanding the population of SBHBs near coalescence in the local
universe can provide insight into the population at high redshift
where theoretical models suggest most mergers take place
\citep{svh07}. In practice however, the search for close binaries and
their post-coalescence counterparts based on their EM signatures has
proven to be a difficult task. There is only a handful of observed
bound SBHB candidates and no confirmed cases at all.  Thus, our
understanding of the accretion physics and associated observational
signatures is poor.  Given the challenges associated with detection of
SBHBs and recoiling SBHs it would be valuable to make predictions of
related observational signatures using hydrodynamical simulations.
While some steps have been taken in that direction, more development
is needed in the future in order to calculate multi-band observational
signatures of these objects.

Our pilot study implies that the period and perhaps the mass ratio of
the binary can be measured from well-sampled, long-term X-ray and
optical light curves and H$\alpha$ profile sequences that have been
followed for at least a few revolutions of the binary. In order to
achieve an efficient SBHB search and avoid time-consuming
observational followups of non-SBHB candidates (false positives) it is
important to utilize all available complementary signatures that may
help to break this degeneracy. Given the location of selected SBHB and
recoiling SBH candidates on the sky, the best prospect for detection
is with radio interferometers, if both black holes have associated
radio sources.  More immediate tests of existence and coalescence of
binaries may be possible in the future in synergy with GW
observations, once the {\it Laser Interferometer Space Antenna} become
operational.

We would like to thank the anonymous referee for insightful comments
and suggestions. TB thanks the UMCP-Astronomy Center for Theory and
Computation Prize Fellowship program for support.

\end{document}